\newif\ifarxiv
\arxivtrue % Comment to remove appendix, and uncomment to compile the appendix

\newif\ifnotarxiv
\ifarxiv\notarxivfalse\else\notarxivtrue\fi

\newif\iftodo
\todotrue

\newif\ifanonymous
\ifarxiv

\documentclass[11pt, english]{article}
\usepackage[a4paper, total={7in, 10in}]{geometry}
\usepackage[T1]{fontenc}
\usepackage{babel}

\usepackage{authblk}

\usepackage{hyperref}
\hypersetup{
    colorlinks=true,
    linkcolor=blue,
    filecolor=magenta, 
    urlcolor=cyan,
    pdftitle={Asymptotically Optimal Tests for One- and Two-Sample Problems},
    pdfpagemode=FullScreen,
}

\else

\documentclass[conference]{IEEEtran}
\IEEEoverridecommandlockouts

\usepackage{url}

\fi

\usepackage{cite}
\usepackage{amsmath,amssymb,amsfonts,amsthm}
\usepackage{mathtools}
\usepackage{algorithmic}
\usepackage{graphicx}
\usepackage{textcomp}
\usepackage[dvipsnames]{xcolor}
\usepackage[nameinlink]{cleveref}
\usepackage{aricksMacros}

\usepackage{soul}
\usepackage{tcolorbox}

\ifarxiv
\date{}
\fi

\ifanonymous
    \ifarxiv
        \author{Anonymous Authors}
    \else

        \author{\IEEEauthorblockN{Anonymous Authors}}

    \fi
\else

    \ifarxiv
        \usepackage{authblk} % Helping with the author block stuff

        \author[1]{Arick Grootveld\thanks{Corresponding author: aegrootv@syr.edu}}
        \author[1]{Biao Chen}
        \author[1]{Venkata Gandikota}

        \affil[1]{Syracuse University}
        
    \else 

        \author{\IEEEauthorblockN{Arick Grootveld}
        \IEEEauthorblockA{
        \textit{Syracuse University}\\
        aegrootv@syr.edu}
        \and
        \IEEEauthorblockN{Biao Chen}
        \IEEEauthorblockA{
        \textit{Syracuse University}\\
        bichen@syr.edu}
        \and
        \IEEEauthorblockN{Venkata Gandikota}
        \IEEEauthorblockA{
        \textit{Syracuse University}\\
        vsgandik@syr.edu}
        
        }
    
    \fi
\fi

\DeclareMathOperator{\TV}{TV}
\newcommand{\RE}[2]{D\left( #1 \middle\| #2 \right)}

\newcommand{\Pxn}{\hat{P}_{X^n}}
\newcommand{\Pyn}{\hat{P}_{Y^n}}
\newcommand{\Pym}{\hat{P}_{Y^m}}

\newcommand{\pxi}{\hat{P}_{X^n}[i]}
\newcommand{\pyi}{\hat{P}_{Y^m}[i]}

\newcommand{\Xn}{X^n}

\newcommand{\Ym}{Y^m}

\newcommand{\Qxn}{\hat{Q}_{X^n}}
\newcommand{\Qyn}{\hat{Q}_{Y^n}}
\newcommand{\Qym}{\hat{Q}_{Y^m}}
\newcommand{\dm}{\delta_m}

\newcommand{\pmin}{p_{\min}}

\def\BibTeX{{\rm B\kern-.05em{\sc i\kern-.025em b}\kern-.08em
    T\kern-.1667em\lower.7ex\hbox{E}\kern-.125emX}}
\begin{document}

\title{Asymptotically Optimal Tests for One- and Two-Sample Problems}

\maketitle

\begin{abstract}

In this work, we revisit the one- and two-sample testing problems: binary hypothesis testing in which one or both distributions are unknown. For the one-sample test, we provide a more streamlined proof of the asymptotic optimality of Hoeffding's likelihood ratio test, which is equivalent to the threshold test of the relative entropy between the empirical distribution and the nominal distribution. The new proof offers an intuitive interpretation and naturally extends to the two-sample test where we show that a similar form of Hoeffding's test, namely a threshold test of the relative entropy between the two empirical distributions is also asymptotically optimal. A strong converse for the two-sample test is also obtained. 
% \ifnotarxiv(``THIS PAPER IS ELIGIBLE FOR THE STUDENT PAPER AWARD''. )\fi
\end{abstract}

\ifnotarxiv
\fi

\section{Introduction}
\label{section:Introduction}
%%%%%%%%%%%%%%%%%%%%%%%%%%%%%%%%%%%%%%%%%%%%%%%%%%
We consider binary hypothesis testing in which one or both distributions are unknown. Specifically, we revisit one- and two-sample testing and study asymptotically optimal schemes for both cases. The one-sample test, also known as goodness of fit test, is to determine if a given sequence of samples follows a nominal distribution. More precisely, let $X^n\sim Q$, the one-sample test is the following binary hypothesis test 
% \begin{equation}
%     H_0: Q=P \quad \mbox{ vs } \quad  H_1: Q\neq P, \label{eq:UHT}
% \end{equation} 
\begin{equation}
    %\label{equation:OneSampleProblem}
    \label{eq:UHT}
    \begin{split}
        H_0: Q = P,\\
        H_1: Q \neq P,
    \end{split}
\end{equation}
where $P$ is the known nominal distribution and $Q$ is unknown under $H_1$. The two-sample test addresses the question of whether two independent sequences follow the same unknown distributions. Given $X^n\sim P$ and $Y^n\sim Q$ with both $P$ and $Q$ unknown, the two-sample test is a binary hypothesis test in exactly the same form as (\ref{eq:UHT}).

%In this work we consider binary universal hypothesis testing problems, where we are asked to distinguish between $H_0: P = Q$ and $H_1: P \neq Q$. There are two special cases of universal problems that we are concerned with, one-sample and two-sample problems. In the one-sample setting we assume complete knowledge of $P$, while our information on $Q$ only comes through samples. In the two-sample setting information about both $P$ and $Q$ is limited to samples from the distributions. 

% In information theory and statistics, a form of comparison between hypothesis testing algorithms is the rate of the asymptotic probability of error. Because of the tradeoff between the types of error, it is common to evaluate the type II error rate when the type I error is constrained, which is known as the Neyman-Pearson setting \cite{neyman1933}. Stein's lemma \cite{cover2006elements} is a celebrated result here, as it connects the Kullback-Leibler divergence (KLD) to the optimal error rate for simple binary hypothesis. 

% The performance of hypothesis testing algorithms can be characterized by the error rates for the possible outcomes. A celebrated result in hypothesis testing is Stein's lemma \cite{cover2006elements}, which shows that under a constraint on the type I errror, the optimal type II error is exactly the KLD. Similarly, 

The one-sample test, often referred to as the universal hypothesis testing in the statistical literature, was formulated and solved in Hoeffding's seminal work \cite{hoeffding1965asymptotically}. Hoeffding showed that the so-called likelihood ratio test achieves the optimal type II error exponent, which is the relative entropy between the nominal distribution and the unknown true distribution under $H_1$. Hoeffding's test is a threshold test of the likelihood ratio between the empirical distribution and the nominal distribution. The test statistic reduces to the simple form of the relative entropy between the empirical distribution and the nominal distribution. Through the use of a properly controlled vanishing threshold, Hoeffding showed that the test is asymptotically optimal in that it achieves the optimal type II error exponent subject to a level constraint on the type I error. %  and the nomi, where he considered tests for multinomial distributions using the sample empirical distribution. Hoeffding's work focused on error rates, as a means of comparison between tests. 
Hoeffding's work inspired a large body of follow-up research, including attempts to extend the solution to continuous distributions \cite{zeitouni2002universal,yang2018robust}. 
A second order analysis, which describes finite sample behavior of the error, was carried out for one-sample tests \cite{harsha2025second}.

% \colorbox{yellow}{} 

% and ... \cite{harsha2025second} studied the second-order asymptotics of empirical divergence tests including the so called Hoeffding test, which computes the Kullback Leibler divergence between the empirical and known distributions. 

The two-sample test, also called homogeneity testing \cite{kanamori2011f} or out-of-distribution sample detection \cite{hendrycks2016baseline}, has found applications in various fields \cite{borgwardt2006integrating, lopez2016revisiting}. While the two-sample test has been studied in the statistical literature, it has attracted significant attention recently in the machine learning community thanks to the discovery of the maximum mean discrepancy (MMD) test \cite{gretton2006kernel} that has shown promising performance across various datasets. This has inspired extensive follow-up research; and it was recently established that MMD with suitable kernels was indeed asymptotically optimal in that it achieves the optimal type II error exponent \cite{zhu2019universal, zhu2021asymptotically} when subject to a level constraint on the type I error. A closely related formulation is the Gutman classification problem\cite{gutman1989asymptotically}, where training sequences $X^n \sim P_1$ and $Y^n \sim P_2$ are given, along with a test sequence $Z^m \sim Q$. The goal is to determine if $Q = P_1$ or $Q = P_2$, using only the empirical samples. In \cite{zhou2020second} a second order asymptotic analysis was performed for the Gutman classification problem. Concurrent to this work, a second order analysis of two-sample hypothesis testing was executed in \cite{harsha2026second}.

%%%%%%%%%%%%%%%%%%%%%%%%%%%%%%%%%%%%%%%%%%%%%%%%%%

In this work, we revisit both the one-sample and two-sample problems for distributions with finite support. 
A more streamlined achievability proof of Hoeffding's likelihood ratio test is provided that also offers a clean interpretation. The proof, as well as the test statistic itself, naturally extends to the two-sample test. More precisely, the threshold test of the relative entropy between the two empirical distributions is shown to achieve the optimal error exponent for the two-sample test, which coincides with Renyi-divergence with $\alpha=\frac{1}{2}$ when the two sequences are of identical length. 
Finally, a strong converse for the two-sample test is obtained, i.e., any test whose type II error probability decays faster than the optimal type II error exponent will have its type I error probability approaching $1$.

\section{Problem Formulation}
\label{section:ProblemFormulation}
\subsection{Notation and Preliminaries}
This section collects all the notation and preliminary lemmas used in the paper.

Take $[d] = \{1,2, \dots, d\}$, $\log$ denotes base 2 logarithm, $\ln$ denotes logarithm base $e$, and we use $\supp(P)$ to describe the support of a probability distribution $P$. For two probability distributions $P,Q$ , $P << Q$ is short hand for $\supp(P) \subseteq \supp(Q)$. We use $\Xn = (X_1, \dots, X_n)$ to refer to a collection of $n$ samples. We say $\Xn \sim Q$ to denote $(X_1, \dots, X_n)$ being independent and identically distributed samples taken from the distribution $Q$. Let $\Pcal^d$ be the probability simplex over $d$ elements such that 
\begin{equation}
    \Pcal^d = \left\{P = (p_1, \dots, p_d) \in [0,1]^d: \sum_{i=1}^d p_i = 1\right\}.
\end{equation}
For $P \in \Pcal^d$, we use $p_{\min}$ to denote the minimum non-zero probability in $P$. 
Denote the total variation distance between distributions $P, Q \in \Pcal^d$ by 
\begin{flalign}
    &\TV(P, Q) = \frac{1}{2} \oneNorm{P - Q},
\end{flalign}
and the Kullback-Leibler divergence (KLD) 
{ 
\begin{flalign} 
    &\RE{P}{Q} = \sum_{i=1}^d p_i \log \frac{p_i}{q_i},
\end{flalign}
}
when $P << Q$, and $+\infty$ otherwise. The $\alpha$-Renyi-relative divergence for $\alpha \in (0,1)$ is defined as 
{
\begin{equation}
    D_{\alpha}(P\| Q) = 
        \frac{1}{\alpha-1} \log \left(\sum_{i=1}^d p_i^{\alpha} q_i^{1-\alpha}\right),
\end{equation}
}
when $\abs{\supp(P)\cap \supp(Q)} \geq 1$, and $+\infty$ otherwise. The Renyi-divergence is symmetric for $\alpha = \frac{1}{2}$. 

For a known distribution $P \in \Pcal^d$ and $c > 0$, we denote the KLD ball %\footnote{The KLD is not a metric, so this is a minor abuse of terminology} 
around $P$ of radius $c$ by 
\begin{align}
    \Bcal(P, c) &= \{F \in \Pcal^d : \RE{F}{P} \leq c\},
\end{align}
and we use $\Bcal^C(P, c)$ to denote its complement in $\Pcal^d$. For $X^n \sim P$ and $Y^m \sim Q$, denote the empirical distributions of these samples as $\Pxn$ and $\Qym$. Throughout this paper, we will make heavy use of the following large deviation result \cite[Theorem 12.4.1]{cover2006elements}
\begin{theorem}[Sanov's Theorem]
    \label{theorem:SanovsTheorem}
    Let $X^n \sim P$ for $P \in \Pcal^d$. Let $E \subset \Pcal^d_n$. Then 
    \begin{equation}
        \Prob[\Pxn \in E] \leq (n+1)^d 2^{-n \RE{F^*}{P}},
    \end{equation}
    where $F^* = \argmin_{F \in E} \RE{F}{P}$. 
    
    % Additionally, if $E$ is the closure of its interior, we have 
    % \begin{equation}
    %     \frac{1}{n} \log P[E] =
    % \end{equation}
\end{theorem}

%We will need a way to compare relative entropies of certain distributions. As such we will make use of what is essentially
For distributions with finite support, the KLD satisfies the following continuity property \cite[Corollary 5.9]{bluhm2023continuity}. 
\begin{lemma}
    \label{lemma:RETransferInequality}
    Suppose $A,B,C \in \Pcal^d$, $A,B << C$, and $\oneNorm{A - B} \leq \eps$. Then 
    \begin{align}
        \abs{\RE{A}{C} - \RE{B}{C}} &\leq 2 \log\left(\frac{1}{c_{\min}}\right) \eps + \frac{2\sqrt{2}}{\ln(2)} \sqrt{\eps}\\
        &\triangleq g(C, \eps),
    \end{align}
    where $c_{\min}$ is the minimum non-zero probability in $C$. 
\end{lemma}

Of note, their result is described for quantum states, while to the best of our knowledge this specific bound had not been shown classically. % prior to the work. 
The only change made here is that for convenience we bound the binary entropy function in their expression using the inequality $$h(x) = -x \ln x - (1-x) \ln(1-x) \leq 2 \sqrt{x (1-x)}.$$ For $\eps$ small enough, $g(C,\eps) = O\left(\sqrt{\eps}\right)$. 

For samples $X^n \sim Q$, a simple binary hypothesis test is to distinguish between the two hypotheses 
\begin{equation}
    \begin{split}
        H_0: Q = P_0,\\
        H_1: Q = P_1,
    \end{split}
\end{equation}
where $P_0, P_1$ are {\em known} distributions. Any test can be described in terms of a function $A_n: [d]^n \to [0,1]$, % that associates $X^n$ to a probability, 
so that for a given $\Xn \in [d]^n$ the test declares $\hat{H}_0$ (accepts $H_0$) with probability $A_n(\Xn)$, and declares $\hat{H}_1$ (rejects $H_0$) with probability $1 - A_n(\Xn)$. 
This definition accommodates deterministic and randomized tests. For a given test $A_n$, denote by $\alpha(A_n) = \Proba{\Hat{H}_1 | H_0}$ the type I error probability and  $\beta(A_n) = \Proba{\hat{H}_0 | H_1}$ the type II error probability. Under the Neyman-Pearson framework, one attempts to minimize $\beta(A_n)$ among all tests that satisfy $\alpha(A_n) \leq \eps$ for some $\eps \in (0,1)$. %, and we consider tests with $\alpha(A_n) \leq \eps$ while attempting to minimize $\beta(A_n)$. 
The following is a well known result for simple hypothesis testing \cite{cover2006elements}. 
%In the simple binary hypothesis testing setting there is a well known result \cite{cover2006elements}: 
\begin{theorem}[Chernoff-Stein Lemma]
    \label{theorem:SteinsLemma}

    For a given $\eps \in (0,1)$ let $\beta_{n}^*(\eps) = \inf_{A_n}\left\{ \beta(A_n): \alpha(A_n) < \eps \right\}$.
    % \begin{equation}
    %     \beta_{n}^*(\eps) = \inf_{A_n}\left\{ \beta(A_n): \alpha(A_n) < \eps \right\}
    % \end{equation}
    % be the best type II error among tests with type I error bounded by $\eps$. 
    We have 
    \begin{equation}
        \lim_{n \to \infty} -\frac{1}{n} \log \beta^*_{n}(\eps) = \RE{P_0}{P_1}.
    \end{equation}
\end{theorem}
Stein's lemma specifies the optimal type II error exponent for the simple hypothesis testing problem when both distributions are known. Naturally, it serves as an upper bound on the type II error exponent for the one-sample test when the distribution under the alternative hypothesis is unknown.

\subsection{One-Sample Problem}
\label{subsection:PF_OS}

For the one-sample test, we are given a known nominal distribution $P$ and samples $\Xn \sim Q$, the one-sample test is specified in (\ref{eq:UHT}) % and we want to  \sim Q$. The two hypothesis are the following: 
% \begin{equation}
%     \label{equation:OneSampleProblem}
%     \begin{split}
%         H_0: Q = P\\
%         H_1: Q \neq P
%     \end{split}
% \end{equation}
and illustrated in \Cref{fig:OS_ProblemDiagram}.

\ifarxiv
\begin{figure}[htb]
    \centering
    \includegraphics[width=0.8\linewidth]{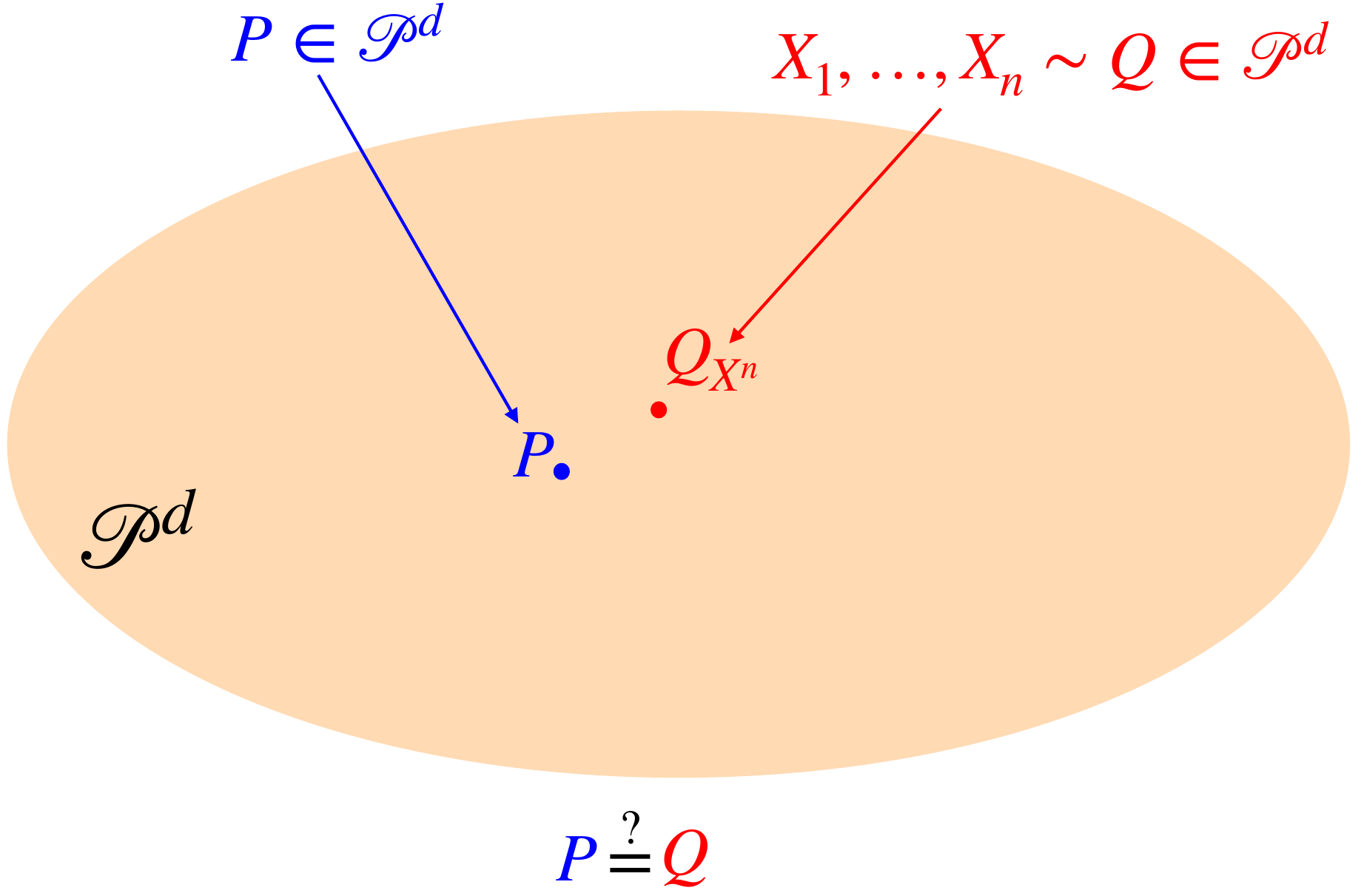}
    \caption{In the one-sample problem we know the nominal distribution $P$, and all information about $Q$ is taken from the sample $X^n$}
    \label{fig:OS_ProblemDiagram}
\end{figure}
\else
\begin{figure}
    \centering
    \includegraphics[width=0.8\linewidth, trim=0 10 0 0]{OS_Diagram.pdf}
    \caption{In the one-sample problem we know the nominal distribution $P$, and all information about $Q$ is taken from the sample $X^n$}
    \label{fig:OS_ProblemDiagram}
\end{figure}
\fi

Hoeffding \cite{hoeffding1965asymptotically} constructed the following test
\begin{equation}
    \label{equation:HoeffdingOSTest}
    \hat{H} = \begin{cases}
        H_0&: \Qxn \in \Bcal(P, c_n),\\
        H_1&: \Qxn \in \Bcal^C(P, c_n),
    \end{cases}
    % \hat{H}=\left\{\begin{split}
    %     H_0&: \Qxn \in \Bcal(P, c_n),\\
    %     H_1&: \Qxn \in \Bcal^C(P, c_n),
    % \end{split} \right.
\end{equation}
% where $c_n n^2 \to \infty$, and $c_n \to 0$. 
Hoeffding refers to this as the likelihood ratio test, as the test statistic,  $D(\Qxn\| P)$, is equivalent to the log likelihood ratio between the empirical distribution and the nominal distribution. He showed that if $c_n \to 0$, this likelihood ratio test, herein denoted by $T_{\text{Hoff},n}$, has the following probabilities: % of type I and II error:
\begin{align}
    \alpha(T_{\text{Hoff},n}) &= 2^{- n c_n + O(\log n)},\\
    \beta(T_{\text{Hoff},n}) &= 2^{-n \RE{P}{Q} + O(\log n)},
\end{align}

Taking $c_n = O\left(\frac{\log n}{n}\right)$ ensures that $\alpha(T_{\text{Hoff},n}) \leq \eps$ for sufficiently large $n$, and 
\begin{equation}
    \label{equation:HoeffdingAchievability}
    \liminf_{n \to \infty} -\frac{1}{n} \log \beta(T_{\text{Hoff},n}) \geq \RE{P}{Q}.
\end{equation}
Combined with \Cref{theorem:SteinsLemma}, this establishes the fact that the test $T_{\text{Hoff},n}$ is asymptotically optimal with type II error exponent $\RE{P}{Q}$. 

%{\color{red} Needs some revision} This is a special case of the more general result by Hoeffding, which is proved through a precise treatment of the error probabilities. However, if we are only concerned with the asymptotics, then a more streamlined proof should be possible. Indeed, in \Cref{subsection:Results_OS} we use modern tools of information theory to prove the one-sample result, which leads to a shorter proof and some insights into the two-sample problem. 

\subsection{Two-Sample Problem}
\label{subsection:PF_TS}

In the two-sample problem we are given samples $\Xn \sim P$ and $\Ym \sim Q$, with both $P$ and $Q$ unknown. The task is to determine if both samples are generated from the same distribution, and the hypothesis test is in exactly the same form as (\ref{eq:UHT}). Here we allow $n \neq m$, but we assume that $\lim_{n,m \to \infty}\frac{n}{n + m} =  w \in \left[\frac{1}{2}, 1\right]$, representing potentially different sample rates for the two distributions. We make the assumption that $n > m$ without loss of generality, as we can simply exchange $X^n$ and $Y^m$ in our definition. 
% The two-sample test involves samples $\Xn \sim P$, and $\Ym \sim Q$, with both $P$ and $Q$ unknown. The task is to determine if both samples are generated from the same distribution, and the hypothesis test is again in exactly the same form as (\ref{eq:UHT}). 
%and our goal is to distinguish between the hypothesis 
% \begin{equation}
%     \begin{split}
%         H_0: Q = P\\
%         H_1: Q \neq P
%     \end{split}
% \end{equation}
\Cref{fig:TS_ProblemDiagram} gives a visual description of this problem. 

\ifarxiv
\begin{figure}[htb]
    \centering
    \includegraphics[width=0.8\linewidth]{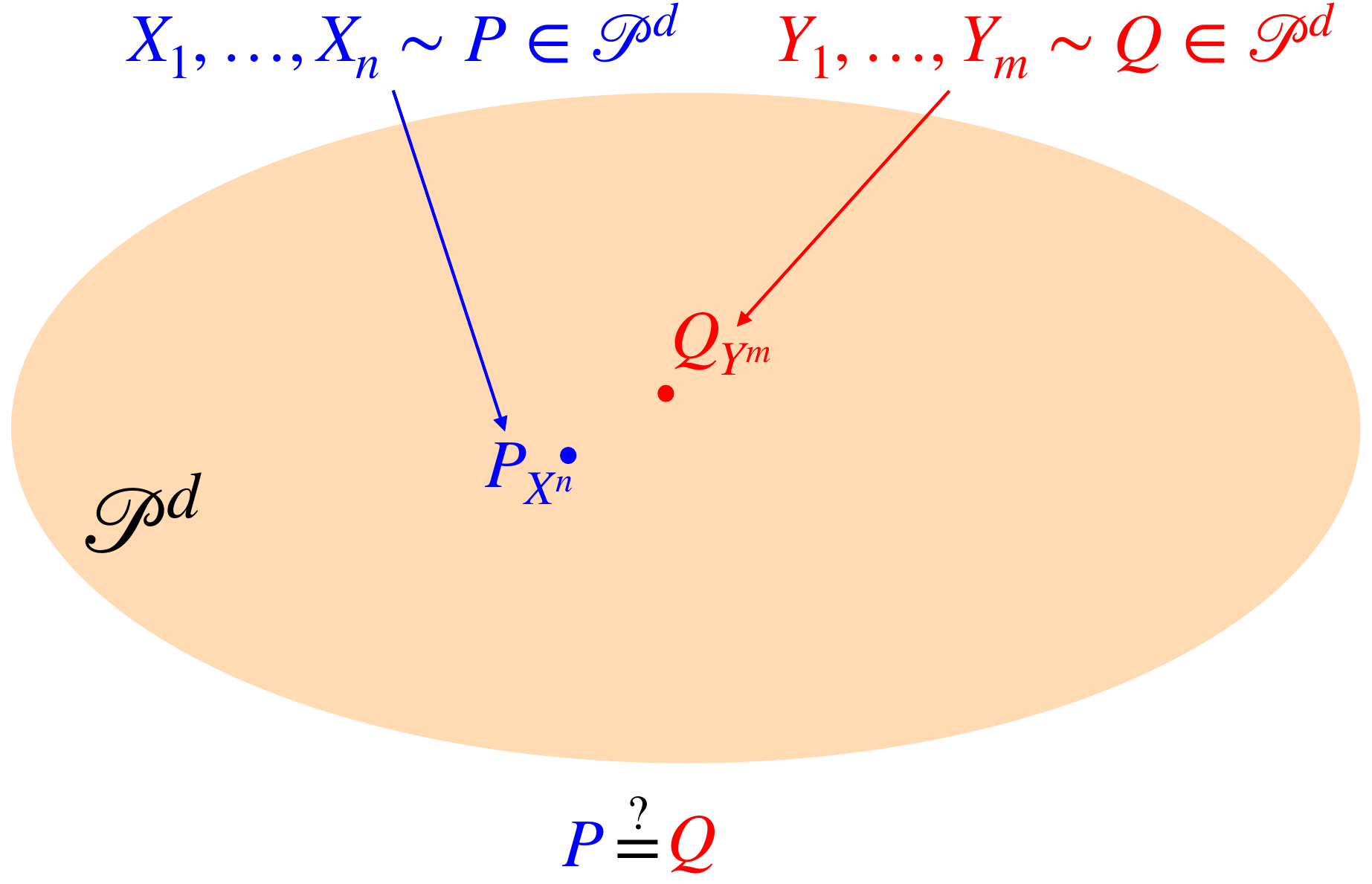}
    \caption{In the two-sample problem we have no knowledge of either true distribution, with all information about $P$ and $Q$ coming from the two samples $X^n$ and $Y^m$}
    \label{fig:TS_ProblemDiagram}
\end{figure}
\else
\begin{figure}
    \centering
    \includegraphics[width=0.8\linewidth, trim=0 10 0 0]{TS_Diagram.pdf}
    \caption{In the two-sample problem we have no knowledge of either true distribution, with all information about $P$ and $Q$ taken from the two samples $X^n$ and $Y^m$}
    \label{fig:TS_ProblemDiagram}
\end{figure}
\fi

For a given $n,m$, the type II error rate is
\begin{equation}
    \beta_{n,m}^*(\eps) = \inf_{A_{n,m}} \{\beta(A_{n,m}): \alpha(A_{n,m}) \leq \eps\}
\end{equation}
The optimal type II error exponent of the two-sample problem was first established in \cite{zhu2019universal, zhu2021asymptotically} to be
{ \small
\begin{equation}
    \label{equation:TS_OptimalRate}
    \begin{split}
        \lim_{\substack{n,m \to \infty}} \frac{-1}{n+m} &\log \beta_{n,m}^*(\eps) 
        \\=  \inf_{F \in \mathcal{P}^d} &w \RE{F}{P} + (1-w)\RE{F}{Q}.
    \end{split}
\end{equation}
}
The achievable scheme proposed in \cite{zhu2019universal, zhu2021asymptotically} relies on metrics that metrize weak convergence. An extended Sanov's theorem, a large deviation result developed in \cite{zhu2019universal, zhu2021asymptotically} involving two independent sequences, was then used to show that the threshold test of any such metric between the two empirical distributions achieves the optimal error exponent for the two-sample setting for the general case, i.e., distributions on Polish, locally compact Hausdorff space such as $R^d$. In contrast, by focusing on the simple case of finite support distributions, we show that direct extension of Hoeffding's test to the two-sample setting is asymptotically optimal. 
%In fact, they addressed a more general problem in two ways. First, their formulation allowed the rate of samples from $P$ and $Q$ to differ, which resulted in a weighted version of what we describe in \Cref{align:TS_OptimalRate}. 

%Here we only address a specialized setting of the previously considered two-sample problem, and in return we show achievability with a test that is simpler, both conceptually and computationally. Additionally we provide a strong converse result for our setting, which was not shown in previous works. Finally, we note that the optimal error exponent is equivalent up to scaling to the Renyi-divergence. 

\section{Results}
\label{section:Results}

\subsection{One Sample Test}
\label{subsection:Results_OS}

Hoeffding's test \cite{hoeffding1965asymptotically} controls the type I error through a vanishing threshold. As the threshold decreases, i.e., the radius of the KLD ball shrinks, the acceptance region, defined by $\Bcal(P,c_n)$ shrinks around $P$. Exploiting the continuity property of KLD in \Cref{lemma:RETransferInequality} for finite alphabet distributions, one can show that the distributions within the shrinking KLD ball ``look similar'' from the perspective of other distributions outside of the ball, i.e., 
for any $Q \in \Pcal^d$ with $Q \neq P$, when $n$ is large enough, we get 
\begin{equation}
    \RE{F}{Q} \approx \RE{P}{Q}, \quad \forall F \in \Bcal(P, c_n).
\end{equation}
This, when combined with Sanov's theorem (\Cref{theorem:SanovsTheorem}), shows the optimality of Hoeffding's test.

\begin{theorem}
    \label{theorem:OS_AchievabilityResult}
    For the test described in \Cref{equation:HoeffdingOSTest} and $c_n = \Theta\left(\frac{\log n}{n}\right)$, we have, for $n$ sufficiently large, 
    \begin{equation}
        \label{equation:OSRes_TypeIError}
        \alpha(A_n) \leq \eps.
    \end{equation}
    
    Further, if $\RE{P}{Q} = +\infty$, then there exists $N > 0$ such that for any $n \geq N$, % we have 
    \begin{equation}
        \label{equation:OSRes_RelEntInfinite}
        \beta(A_n) = 0.
    \end{equation}

    If $\RE{P}{Q} < +\infty$, then instead we get 
    \begin{equation}
        \label{equation:OsRes_RelEntFinite}
        \lim_{n \to \infty} -\frac{1}{n} \log \beta(A_n) = \RE{P}{Q}.
    \end{equation}
\end{theorem}
The proof can be found in {\ifarxiv \Cref{section:Appendix_OS_Achievability}
\else 
\Cref{subsection:Appendix_OS_Achievability}
\fi}. 

While the optimality property was originally established in \cite{hoeffding1965asymptotically}, our proof, which offers a simple geometric interpretation, allows one to directly construct a test for the two-sample problem with the desired optimality.

\subsection{Two Sample Test}
\label{subsection:Results_TS}

Recall that $\Pxn$ and $\Qym$ denote respectively the empirical distributions of $\Xn$ and $\Ym$. The one-sample test uses the KLD ball around the nominal distribution with a vanishing radius as the acceptance region. The rate with which it vanishes is controlled in a manner to ensure type I error constraint is satisfied, i.e., the probability that the empirical distributions fall within the KLD ball should be sufficiently large. The same geometric interpretation can be used in constructing a test that works for the two-sample setting. As the sample size increases, the empirical distributions will concentrate around the true distribution, i.e., $\Pxn$ will likely fall within the KLD ball centered at $P$ whereas $\Qym$ will likely fall within the KLD ball centered at $Q$. If $P=Q$, then the KLD ball will overlap and one can resort to Bonferroni inequality to argue that $\Pxn$ will concentrate around the KLD ball centered at $\Qym$ and vice versa. This motivates the following test for the two-sample problem.
\begin{equation}
    \label{align:TS_TestAlg}
    \hat{H} = \begin{cases}
        H_0 &: \RE{\Pxn}{\Qym} \leq c_m,\\
        H_1 &: \RE{\Pxn}{\Qym} > c_m,
    \end{cases}
\end{equation}
where $c_m = \Theta(m^{-1/2})$ is chosen to satisfy a type I error constraint\footnote{Our choice of $c_m = \Theta(m^{-1/2})$ is arbitrary, the result holds for  $c_m = m^{-1 +\gamma}$ for any $\gamma \in (0,1)$.}. 
The following theorem states that this test achieves the optimal type II error exponent. 
\begin{theorem}[Achievability]
    \label{theorem:TS_Achievability}
    Given $\eps \in (0,1)$ the constraint on the type I error, using the test described in \Cref{align:TS_TestAlg} as $A_{n,m}$, we have, $\forall P \in \Pcal^d$,  
    \begin{equation}
        \label{equation:TS_TypeI}
        \limsup_{n,m \to \infty} \Proba{(\Pxn, \Pym) \in A_{n,m}^C | \Xn, \Ym \sim P } \leq \eps.
    \end{equation}

    And for $P, Q \in \Pcal^d$ with $\supp(P) \cap \supp(Q) = \emptyset$, 
    \begin{equation}
        \label{equation:TS_TypeII_DiffSupp}
        \Proba{(\Pxn, \Qym) \in A_{n,m} | (\Xn, \Ym) \sim (P,Q)} = 0.
    \end{equation}

    Finally, for $P,Q \in \Pcal^d$ with $\supp(P) \cap \supp(Q) \neq \emptyset$, we get
    { \small
    \begin{align}
        \label{align:TS_TypeII_SharingSupp}
        \begin{split}
            \liminf_{n,m \to \infty} -\frac{1}{n+m} &\log \Proba{(\Pxn, \Qym) \in A_{n,m} | (\Xn, \Ym) \sim (P,Q)} \\
            &\geq w\RE{F^*}{P} + (1-w)\RE{F^*}{Q}.
        \end{split}
    \end{align}
    }

    where $F^* = \argmin_{F \in \Pcal^d} w\RE{F}{P} + (1-w)\RE{F}{Q}$. 
\end{theorem}
The proof  
can be found in {\ifarxiv
\Cref{section:Appendix_TS_Achievability}
\else
\Cref{subsection:Appendix_TS_Achievability}
\fi
}. 

Notice that the test described in \Cref{align:TS_TestAlg} is not symmetric in $P$ and $Q$, while the problem itself is, both in the problem formulation and in the optimal error exponent. Indeed, the result holds when replacing $\RE{\Pxn}{\Qym}$ with $\RE{\Qym}{\Pxn}$, or the minimum of the two. Using the minimum test statistic yields a better type I error while achieving the same asymptotic type II exponent. 

We now show a converse result for the two-sample problem. 
\begin{theorem}[Converse]
    \label{theorem:TS_Converse}
    Take $T_{n,m}$ to be a sequence of tests and $\eps \in (0,1)$ such that $\forall F \in \Pcal^d$
    \begin{equation}
        \limsup_{n,m\to \infty} \Prob[(\Pxn, \Qym) \in T_{n,m}^C | X^n, Y^m \sim F] \leq \eps.
    \end{equation}
    Then $\forall P, Q \in \Pcal^d$ with $P \neq Q$ and $\supp(P) \cap \supp(Q) \neq \emptyset$,
    \begin{equation}
        \limsup_{n,m \to \infty} -\frac{1}{n+m} \log \Prob[(\Pxn, \Qym) \in T_{n,m}| X^n \sim P, Y^m \sim Q] \leq w D(F^*\|P) + (1-w) D(F^*\|Q).
    \end{equation}
\end{theorem}
{\ifarxiv 
The proof can be found in \Cref{section:Appendix_TS_Converse}
\else
The full proof can be found in the extended version of this work \cite{grootveld2026asymptotically}.
\fi
}

Given $P,Q$, solving for $F^* = (f_k^*)_{k}$ yields 
\begin{equation}
    \label{equation:optimalF}
    f_k^* = \frac{p_k^{w} q_k^{1-w}}{\sum_{k=1}^d p_i^{w} q_i^{1-w}}.
\end{equation}
Notice that $F^*$ is a geometric mixture of the two distributions. It follows that, from \cite[Theorem 30]{van2014renyi}, the optimal type II error exponent for the two-sample test is precisely the $w$ Renyi-relative entropy, i.e.,% it was shown that the Renyi-relative entropy gives an equivalent interpretation, 
\begin{equation}
    \label{equation:RenyiRel_MidPointInterp}
    w\RE{F^*}{P} + (1-w)\RE{F^*}{Q} = (1-w) D_{w}(P \| Q).
\end{equation}

In combination \Cref{theorem:TS_Achievability}, \Cref{theorem:TS_Converse}, and \eqref{equation:RenyiRel_MidPointInterp}, % tell us that
show that the test in \Cref{align:TS_TestAlg} satisfies
{\small
\begin{equation}
    \begin{split}
    \lim_{n,m \to \infty} -\frac{1}{n+m} \log \Prob&\left[(\Pxn, \Qym) \in A_{n,m} | (\Xn, \Ym) \sim (P,Q)\right] \\
    &= (1-w) D_{w}(P\| Q).
    \end{split}
\end{equation}
}

We also show a strong converse for the two-sample problem. 
\begin{theorem}
    \label{theorem:TS_StrongConverseRate}
     Fix a particular $P,Q \in \Pcal^d$ with $\supp(P) \cap \supp(Q) \neq \emptyset$ and $P \neq Q$. For a given test with acceptance region $T_{n,m} \subseteq \Pcal^d_n \times \Pcal_m^d$, assume that  there exists $\delta > 0$ such that  
    {
    \begin{flalign}
        \begin{split}
            &\liminf_{n,m \to \infty} -\frac{1}{n+m} \log \Proba{(\Pxn, \Qym) \in T_{n,m} | (\Xn, \Ym) \sim (P,Q)}\\
            &\geq w\RE{F^*}{P} + (1-w)\RE{F^*}{Q} + \delta.
        \end{split}
    \end{flalign}
    }
    Taking $F^*$ as in \Cref{equation:optimalF}, we find that $\forall \eta \in (0, \delta)$
    \begin{flalign}
        \begin{split}
            &\liminf_{n,m \to \infty} -\frac{1}{n+m} \log \Proba{(\Pxn, \Pym) \in T_{n,m} | X^n, Y^m \sim F^*} \\
            &\geq \sup_{0\leq s < 1-w} \left\{s\eta + (1-s) \left[\psi(w) - w \psi\left(\frac{w-s}{1-s}\right) - (1-w) \psi\left(\frac{w}{1-s}\right)\right] \right\}\\
            &> 0,
        \end{split}
    \end{flalign}
    where $\psi(s) = \log \left(\sum_{i=1}^d p_i^s q_i^{1-s}\right)$. 
\end{theorem}
{\ifarxiv
The proof can be found in \Cref{section:Appendix_TS_StrongConverseRate}.
\else
The full proof can be found in the extended version of this work\cite{grootveld2026asymptotically}.
\fi
}

From \Cref{theorem:TS_StrongConverseRate}, we can see that the probability of accepting the null hypothesis when $X^n, Y^m \sim F$ decays exponentially, giving us a strong converse result. 
\begin{corollary}[Strong Converse]
    \label{corollary:TS_StrongConverse}
    For a test with an acceptance region $T_{n,m}$, if for a particular $P, Q \in \Pcal^d$ with $\supp(P) \cap \supp(Q) \neq \emptyset$ and $P \neq Q$, we have 
    {\small
    \begin{flalign}
        \begin{split}
            &\liminf_{n,m \to \infty} -\frac{1}{n+m} \log \Proba{(\Pxn, \Qym) \in T_{n,m} | (\Xn, \Ym) \sim (P,Q)}\\
            &> w\RE{F^*}{P} + (1-w)\RE{F^*}{Q}.
        \end{split}
    \end{flalign}
    }

    Then
    \begin{align}
        \lim_{n,m \to \infty} \Proba{(\Pxn, \Qym) \not\in T_{n,m} | \Xn, \Ym \sim F^*} = 1. \label{eq:F}
    \end{align} 
\end{corollary}
% {\ifarxiv 
% The proof can be found in \Cref{section:Appendix_TS_StrongConverse}.
% \else
% The full proof can be found in the extended version of this work \cite{grootveld2026asymptotically}.
% \fi
% }

\ifnotarxiv

\section{Proofs}

% \subsection{Proof of \texorpdfstring{\Cref{theorem:OS_AchievabilityResult}}{One Sample Achievability} }
\subsection{Proof of \Cref{theorem:OS_AchievabilityResult}}
\label{subsection:Appendix_OS_Achievability}

\begin{proof}
    To get \Cref{equation:OSRes_TypeIError}, we apply \Cref{theorem:SanovsTheorem}, giving us
    \begin{equation}
        \Proba{\Qxn \in \Bcal^C(P, c_n) | X^n \sim P} \leq (n+1)^d 2^{-n c_n}.
    \end{equation}
    Taking
    \begin{equation}
        c_n = \frac{d \log(n+1)}{n} + \frac{\log(1/\eps)}{n},
    \end{equation}
    gives us the type I error constraint $\Proba{\hat{H}_1 | H_0} \leq \eps$. 

    To see \Cref{equation:OSRes_RelEntInfinite}, we note that the condition is equivalent to $\exists k \in [d]$ such that $p_k > 0$ but $q_k = 0$. A type II error occurs if $\RE{\Qxn}{P} \leq c_n$. Combined with Pinsker's inequality \cite{cover2006elements}, we have
    \begin{equation}
         c_n \geq \frac{1}{2 \ln(2)} \oneNorm{\Qxn - P}^2,
    \end{equation}
    Then for $c_n < \frac{p_{\min}^2}{ 2 \ln(2)}$ we must have $\Qxn[k] > 0$, however since $Q[k] = 0$ this is a probability 0 event. 

   % As discussed at the end of \Cref{subsection:PF_OS}, due to the converse result in 
   From \Cref{theorem:SteinsLemma}, the $D(P\|Q)$ is the optimal type II error exponent for the simple hypothesis test and is thus an upper bound for the one-sample test as $Q$ is unknown. To show \Cref{equation:OsRes_RelEntFinite}, we need only give an achievability proof. Furthermore, by \Cref{theorem:SanovsTheorem} it is sufficient to show that for any $\delta > 0$ there is $N_{\delta} > 0$ such that
    \begin{equation}
        \label{equation:Satisfiable_OS}
        \min_{\Qxn \in \Bcal(P, c_n)} \RE{\Qxn}{Q} \geq \RE{P}{Q} - \delta, \quad \forall n \geq N_{\delta}.
    \end{equation}

    Now, note that for any $F << Q$ with $F \in \Bcal(P, c_n)$, we have 
    \begin{align}
        c_n \geq \RE{F}{P} &\geq \frac{1}{2\ln 2} \oneNorm{F-P}^2\\
        \implies \oneNorm{F - P} &\leq \sqrt{c_n 2 \ln 2}.
    \end{align}

    Invoking \Cref{lemma:RETransferInequality}, we have 
    \begin{equation}
        \abs{\RE{F}{Q} - \RE{P}{Q}} \leq g\left(Q, \sqrt{c_n 2 \ln 2}\right).
    \end{equation}

    However, since $g(Q,x)$ is continuous in $x$ and $g(Q,0) = 0$ we have $\exists U_{\delta} > 0$ such that $\forall \tau \in [0,U_{\delta}]$, $g(Q, \tau) \leq \delta$. But since $c_n \searrow 0$, $\exists N_{\delta}> 0$ such that % which has 
    \begin{equation}
        \sqrt{c_n 2 \ln 2} \in [0, U_{\delta}] , \quad \forall n \geq N_{\delta}.
    \end{equation}

    Therefore, we have 
    \begin{equation}
        \abs{\RE{F}{Q} - \RE{P}{Q}} \leq \delta,
    \end{equation}
    and since this holds for all $F \in \Bcal(P,c_n)$, we have \Cref{equation:Satisfiable_OS}. 
\end{proof}

% \subsection{Proof of \Cref{lemma:SufficientStatistics}}
% \label{subsection:Appendix_TS_SufficientStats}

% \begin{proof}
%     Since 
%     \begin{equation}
%         \Prob_{(P,Q)}\left[(\Xn, \Yn)\right] = \prod_{i=1}^d p_i^{n \cdot \Pxn[i]} \prod_{j=1}^d q_j^{n \cdot \Qyn[j]},
%     \end{equation}

%     by Fisher's factorization theorem \cite{cover2006elements} we have $(\Pxn, \Qyn)$ being a sufficient statistic for $(P,Q)$. And sufficiency for $(P,Q)$ implies sufficiency for $H_0: P=Q$, $H_1: P\neq Q$ since this is a function of $(P,Q)$ only. 
% \end{proof}

% \subsection{Proof of \texorpdfstring{\Cref{theorem:TS_Achievability}}{Two-Sample Achievability} }
\subsection{Proof of \Cref{theorem:TS_Achievability}}
\label{subsection:Appendix_TS_Achievability}

\begin{proof}

    \Cref{equation:TS_TypeII_DiffSupp} is obvious from the fact that $c_{n} < \infty$. 

    Now we analyze \Cref{equation:TS_TypeI}. Let 
    \begin{align}
        \dm &= \frac{1 + \log(1/\eps)}{m} + \frac{d\log m}{m} = \Theta\left(\frac{\log m}{m}\right),\\
        c_{m} &= \dm + m^{-1/2} = \Theta(m^{-1/2}).
    \end{align}
     Now, 
    { \footnotesize
    \begin{flalign}
        &\Proba{\hat{H}_1 | H_0}\\
        &=\Prob\left[\RE{\Pxn}{\Pym} \geq c_{m} \middle| \Pxn \in \Bcal(P, \dm)\right]\Proba{\Pxn \in \Bcal(P, \dm)}\\
        &+ \Prob\left[\RE{\Pxn}{\Pym} \geq c_{m} \middle| \Pxn \in \Bcal^C(P, \dm)\right]\Proba{\Pxn \in \Bcal^C(P, \dm)}\\
        &\leq \Proba{\RE{\Pxn}{\Pym} \geq c_{m} | \Pxn \in \Bcal(P, \dm)} + (n+1)^d 2^{-n\dm}\\
        &\leq \Proba{\RE{\Pxn}{\Pym} \geq c_{m} | \Pxn, \Pyn \in \Bcal(P, \dm)} + 2(n+1)^d 2^{-n\dm}.
    \end{flalign}
    }
    Note that $2 (n+1)^d 2^{-n \dm} \leq \eps$ for $n$ sufficiently large. 

    Assume that $\RE{\Pxn}{P}, \RE{\Pym}{P} \leq \delta_{n,m}$. 
    % Then using Pinsker's inequality again, we have 
    % \begin{equation}
    %     \oneNorm{\Pxn - P}, \oneNorm{\Pyn - P} \leq \sqrt{\delta_n 2 \ln 2}
    % \end{equation}
    Therefore, $\exists N_1 > 0$ such that for all $ n \geq N_1$ we have 
    \begin{align}
        \oneNorm{\Pym - P} &\leq \frac{\pmin}{2}\\
        \implies \pyi &\geq \frac{\pmin}{2}, \quad \forall i \in [d].
    \end{align}
    
    Now, using the fact that $\ln(x) \leq x - 1$ for $x > 0$, we get
    { \small
    \begin{align}
        \RE{\Pxn}{\Pym} &= \sum_{i=1}^d \pxi \log\frac{\pxi}{\pyi}\\
        &\leq \frac{1}{\ln 2}\sum_{i=1}^d \pxi \frac{\pxi - \pyi}{\pyi}\\
        % &= \frac{1}{\ln 2} \sum_{i=1}^d \frac{(\pxi - \pyi)^2}{\pyi}\\
        &\leq \frac{2}{\pmin \ln 2} \oneNorm{\Pxn - \Pym}^2\\
        &\leq \frac{16}{\pmin} \delta_n, 
    \end{align}
    }
    where the last inequality comes from applying Pinsker's inequality to our assumption on $\Pxn, \Pyn$. Since $c_n = \Theta(n^{-1/2})$ while $\delta_{n,m} = \Theta(\frac{\log n}{n})$, $\exists N_2 > 0$ such that $\forall n \geq N_2$ we have 
    \begin{equation}
        \frac{16}{\pmin} \delta_n < c_m
    \end{equation}

    Consequently, for $n \geq \max\{N_1, N_2\}$ we get
    {\small
    \begin{align}
        &\RE{\Pxn}{\Pym} \leq \frac{16}{\pmin}\delta_{n,m} < c_m\\
        \implies &\Prob\left[\left.\RE{\Pxn}{\Pym} > c_n \right| \Pxn, \Pym \in \Bcal(P, \delta_n)\right] = 0,
    \end{align}
    }
    which gives us the type I error bound. 
    %%%%%%%%%%%%%%%%%%%%%%%%%%%%%%%%%%%%%%%%%%%%
    
    Finally, to prove \Cref{align:TS_TypeII_SharingSupp} we use a similar argument to the proof of  \Cref{equation:OsRes_RelEntFinite}. Specifically, in order to have a type II error occur for this test, we must have $\Pxn, \Qyn$ close together. And for any $\eta > 0$ if they are sufficiently close, then by \Cref{lemma:RETransferInequality} they will have 
    \begin{equation}
        \abs{\RE{\Pxn}{P} - \RE{\Qym}{P}} \leq g(P, \eta),
    \end{equation}
    so that 
    { \small
    \begin{align}
        &w\RE{\Pxn}{P} + (1-w)\RE{\Qyn}{Q} \\
        &\leq w\RE{\Qym}{P} + (1-w)\RE{\Qym}{Q} + (1-w)g(P,\eta)\\
        &\leq w\RE{F^*}{P} + (1-w)\RE{F^*}{Q} + g(P,\eta).
    \end{align}
    }

    And since $g(P, \eta) \to 0$ as $\eta \to 0$, and $\eta$ is arbitrary we recover the result.

    The existence and uniqueness of $F^*$ comes from the fact that we are minimizing a convex function over $\Pcal^d$, which is a compact and convex set \cite{boyd2004convex}. 
\end{proof}

% \subsection{Proof of \texorpdfstring{\Cref{theorem:TS_StrongConverse}}{Strong Converse} }
% \subsection{Proof of \Cref{theorem:TS_StrongConverse}}
% \label{subsection:Appendix_TS_StrongConv}

% % To conserve space, we provide only a high level sketch.  

% \begin{proof}

%     % By the theorem statement we can assume that there is some $\eps > 0$ such that 
%     % \begin{flalign}
%     %     \begin{split}
%     %         -&\liminf_{n,m \to \infty} \frac{1}{n+m} \log \Proba{(\Pxn, \Qym) \in T_{n} | (\Xn, \Ym) \sim (P,Q)}\\
%     %         &= \left[w\RE{F^*}{P} + (1-w)\RE{F^*}{Q} + \eps\right].
%     %     \end{split}
%     % \end{flalign}
    
%     % For this $\eps$, we can find $c_{\eps} > 0$ with the property that $\forall A,B \in \Bcal(F^*, c_{\eps})$ 
%     % \begin{equation}
%     %     w\RE{A}{P} + (1-w)\RE{B}{Q} < w\RE{F^*}{P} + (1-w)\RE{F^*}{Q} + \eps.
%     % \end{equation}
%     % The acceptance region $T_{n,m}$ will eventually have no overlap with $\Bcal(F^*, c_{\eps}) \times \Bcal(F^*, c_{\eps})$, since this would violate the assumption on our type II error exponent. 
    
%     % Now, suppose that $X^n, Y^m \sim F^*$, the empirical distributions must converge to the true distribution \cite{van2000asymptotic}, i.e.
%     % \begin{equation}
%     %     \Fxn, \Fyn \overset{\Prob}{\to} F^*.
%     % \end{equation}

%     % Therefore we have
%     % \begin{align*}
%     %     &\lim_{n \to \infty} \Prob[(\Fxn, \Fym) \in \Bcal(F^*, c_{\eps}) \times \Bcal(F^*, c_{\eps})] = 1,
%     % \end{align*}
%     % which bounds the type I error. 

% \end{proof}

\fi

\section{Conclusion}
\label{section:Conclusions}

This paper provided a more streamlined proof of the asymptotic optimality of Hoeffding's likelihood ratio test for the one-sample problem. The test statistic and the proof itself naturally extend to the two-sample test, and we show that a simple test comparing the relative entropy between the two empirical distributions is asymptotically optimal for the two-sample problem. Our matching converse bound shows that this is the asymptotically optimal error exponent for two-sample hypothesis testing. Finally, we established a strong converse for the two-sample test, which gives the following interpretation: any test whose type II error probability vanishes faster than the optimal type II error exponent necessarily has its type I error approaching one exponentially fast. 
%In the one-sample universal setting we describe a new proof of an established result that the KLD between the empirical and given distribution is a a good statistic, in the sense that it achieves the optimal error exponent. In the two-sample setting we prove that the KLD between the empirical distributions is a good statistic in the same sense. Finally, we show a two-sample strong converse, which improves upon the prior known results. 

%One promising avenue of research in the two sample setting is a characterization of the optimal type II error rate when the type I error  decays exponentially fast. A second direction is to extend the strong converse to distributions with continuous support, and regimes with unequal sample sizes. 

\ifnotarxiv
\clearpage
\fi
\bibliographystyle{IEEEtran}
\bibliography{refs.bib}

\ifarxiv

\appendix

\section{Preparatory Results}
\label{section:Appendix_Preliminaries}

We have the following results that we will need: 
\begin{theorem}[Pinsker's Inequality \cite{cover2006elements}]
    \label{theorem:PinskersInequality}
    \begin{equation}
        D(P\|Q) \geq \frac{1}{2\ln(2)} \oneNorm{P - Q}^2
    \end{equation}
\end{theorem}

We also will require a bound on the probability of recovering a particular empirical distribution. 
\begin{theorem}[Theorem 11.1.4 \cite{cover2006elements}]
    \label{theorem:SpecificTypeBound}
    \begin{equation}
        (n+1)^{-d} 2^{-n \RE{\dot{P}}{Q}} \leq \Proba{\Pxn = \dot{P} | X^n \sim Q} \leq 2^{-n \RE{\dot{P}}{Q}}
    \end{equation}
\end{theorem}

In the process of proving our strong converse, we will need the following specialization of Gibbs variational principle \cite[Proposition 4.7]{polyanskiy2025information}.
\begin{theorem}[Gibbs Variational Principle]
    \label{theorem:GibbsVariationalPrinciple}
    For $c_1, \dots, c_d \in \R$ and $Q \in \Pcal^d$,  
    \begin{equation}
        \inf_{P \in \Pcal^d} \left\{ D(P\|Q)  - \sum_{i=1}^d p_i c_i \right\} = - \log \left(\sum_{i=1}^d q_i 2^{c_i}\right)
    \end{equation}
\end{theorem}

\section{Proof of \texorpdfstring{\Cref{theorem:OS_AchievabilityResult}}{One-Sample Achievability}}
\label{section:Appendix_OS_Achievability}

\subsection{Type I Error}

Under the null hypothesis, we will denote $Q_{X^n}$ with $P_{X^n}$ to clarify that $P = Q$ here. Now, using \Cref{theorem:SanovsTheorem} we have 
\begin{align*}
    \Proba{\hat{H}_1 | H_0} &= \Proba{\RE{\Pxn}{P} > c_n}\\
    &\leq (n+1)^d 2^{-n \cdot c_n}
\end{align*}

Then if $\eps, d$ are fixed, we have 
\begin{align}
    c_n &= \frac{d \log(n+1)}{n} + \frac{\log(1/\eps)}{n}\\
    &= \Theta\left(\frac{\log(n)}{n}\right)
\end{align}

which would cause $\Proba{\hat{H}_1 | H_0} \leq \eps$.

\subsection{Type II Error}

%%%%%%%%%%%%%%%%%%%%%%%%%%%%%%%%%%%%%%%%%%%%%%%%%%%
%%%%%%%%%%%%%%%  D(P||Q) = +\infty  %%%%%%%%%%%%%%%
%%%%%%%%%%%%%%%%%%%%%%%%%%%%%%%%%%%%%%%%%%%%%%%%%%%
If $\RE{P}{Q} = +\infty$, then $\exists k \in [d]$ such that $P[k] > 0$, but $Q[k] = 0$. Here we use the notation $F[k] := \Prob_{X \sim F}[X = k]$. 

Take $p_{\min} := \min_{i \in \supp(P)} p_i$. Since $c_n \searrow 0$ as $n\to \infty$, for $n$ large enough we get 
\begin{equation}
    (2 \ln(2))^{1/2} c_n^{1/2} < p_{\min}
\end{equation}

Then if $Q_{X^n} \in B_n(P,c_n)$ 
\begin{align}
    \frac{1}{2\ln(2)} \oneNorm{P - Q_{X^n}}^2 &\leq \RE{Q_{X^n}}{P} \leq c_n\\
    \implies \oneNorm{Q_{X^n} - P} \leq (2\ln(2))^{1/2} c_n^{1/2} < p_{\min}
\end{align}

Therefore, for $n$ large enough, and $Q_{X^n} \in B_n(P, c_n)$, we have $P << Q_{X^n}$. But then $Q_{X^n}[k] > 0$, and since $X^{n} \sim Q$ the probability of seeing such a $Q_{X^n}$ is 0 since $Q[k] = 0$. 

So, if $\RE{P}{Q} = + \infty$, and $n$ large enough, we have 
\begin{equation}
    \Proba{\RE{Q_{X^n}}{P} \leq c_n} = 0
\end{equation}

\vspace{1cm}
%%%%%%%%%%%%%%%%%%%%%%%%%%%%%%%%%%%%%%%%%%%%%%%%%%%
%%%%%%%%%%%%%%%  D(P||Q) < +\infty  %%%%%%%%%%%%%%%
%%%%%%%%%%%%%%%%%%%%%%%%%%%%%%%%%%%%%%%%%%%%%%%%%%%
Now we prove the result when $\RE{P}{Q} < +\infty$. Additionally, without loss of generality we can assume that for all $Q_{X^n} \in \Pcal_n^d$ that  $\RE{Q_{X^n}}{Q} < +\infty$, i.e. $\Qxn << Q$. 

\begin{align}
    \Proba{\hat{H}_0 | H_1} &= \Proba{ \Qxn \in B_n(P, c_n)}\\
    &\leq (n+1)^d 2^{ -n  \ \cdot \min\limits_{Q_{X^n} \in B(P, c_n)} \{\RE{ Q_{X^n} }{Q}\} }
\end{align}

Which means we can focus on bounding
\begin{equation}
    \min\limits_{Q_{X^n} \in B(P,c_n)} \{\RE{Q_{X^n}}{Q}\}
\end{equation}
from below. 

For $Q_{X^n} \in B(P,c_n)$, we have
\begin{align}
    \frac{1}{2\ln(2)} \oneNorm{Q_{X^n} - P}^2 &\leq D(Q_{X^n} \| P) \leq c_n\\
    \implies \oneNorm{Q_{X^n} - P} &\leq (2\ln(2))^{1/2} c_n^{1/2} := r_n
\end{align}

Since $c_n \searrow 0$ we get $r_n \searrow 0$. And by our assumptions, we have $\RE{P}{Q} < +\infty$, and $\RE{Q_{X^n}}{Q} < +\infty$.  Then invoking \Cref{lemma:RETransferInequality}, we have 
\begin{equation}
    \abs{\RE{Q_{X^n}}{Q} - \RE{P}{Q}} \leq g(Q, r_n)
\end{equation}

Now, for $Q$ and $d$ fixed, $g(Q, r_n)$ is monotonically decreasing in $r_n$, and $g(Q,0) = 0$. Therefore $\forall \eps > 0$, $\exists N > 0$ such that $\forall n > N$
\begin{align}
    \abs{\RE{Q_{X^n}}{Q} - \RE{P}{Q}} &\leq \eps\\
    \implies \RE{Q_{X^n}}{Q} &\geq \RE{P}{Q} - \eps\\
    \implies \min_{Q_{X^n} \in B(P,c_n)} \RE{Q_{X^n}}{Q} &\geq D(P\| Q) - \eps
\end{align}

So that for $n$ large enough we have
\begin{align}
    \Proba{\RE{Q_{X^n}}{P} \leq c_n} &\leq (n+1)^d 2^{-n \left(D(P\| Q) - \eps\right)}\\
    \implies \liminf_{n \to \infty} -\frac{1}{n} \log \beta(A_n) &\geq \RE{P}{Q}
\end{align}

The other direction follows directly from the converse in Stein's lemma, see for example \cite{cover2006elements}.

\section{Proof of \texorpdfstring{\Cref{theorem:TS_Achievability}}{Two-Sample Achievability}}
\label{section:Appendix_TS_Achievability}

\subsection{Type I Error}
In this section, we will prove that the test described above will asymptotically achieve a Type I error of $\alpha$. By this we mean that given $\alpha \in (0,1)$ for $n,m$ large enough, the test will have Type I error $\leq \alpha$. Under $H_0$, we will denote the two empirical distributions by $\Pxn$ and $\Pym$, to signify that the underlying distribution for both $X^n$ and $Y^m$ is $P$. 

Now, for any $\delta_{m} > 0$, we have 
\begin{align}
    \Proba{\hat{H}_1| H_0} &= \Proba{\RE{\Pxn}{\Pym} > c_m}\\
    &= \Proba{\RE{\Pxn}{\Pym} > c_m | \Pxn \in B(P, \delta_m)}\Proba{\Pxn \in B(P, \delta_{m})}\\
    & \  + \Proba{\RE{\Pxn}{\Pym} > c_m | \Pxn \in B^C(P, \delta_{m})} \Proba{\Pxn \in B^C(P, \delta_{m})}\\
    &\leq \Proba{\RE{\Pxn}{\Pym} > c_m | \Pxn \in B(P, \delta_{m})} + \Proba{\Pxn \in B^C(P, \delta_{m})}\\
    &\leq \Proba{\RE{\Pxn}{\Pym} > c_m | \Pxn \in B(P, \delta_{m})} + (n+1)^d 2^{-n\delta_{m}}\\
    &\leq \Proba{\RE{\Pxn}{\Pym} > c_m | \Pxn, \Pym \in B(P, \delta_{m})} + (n+1)^d 2^{-n\delta_{m}} + (m+1)^d 2^{-m \delta_{m}}
\end{align}

Since we have $m < n$ by assumption, for $m$ large enough we get 
\begin{align}
    (n+1)^d 2^{-n \delta_{m}} + (m+1)^d 2^{- m \delta_{m}} &\leq 2(m+1)^d 2^{-m \delta_{m}}
\end{align}

Now, if we take 
\begin{align}
    \delta_{m} = \frac{1 + \log(1 / \eps) + d\log(m+1)}{m} = \Theta\left(\frac{\log(m)}{m}\right)
\end{align}
which gives us $2(m+1)^d 2^{-m\dm} \leq \eps$. Further, take\footnote{The result would hold if we took $c_m = \dm + m^{-\gamma}$ for any $\gamma \in (0,1)$} $c_m = \dm + m^{-1/2} = \Theta(m^{-1/2})$. 

Then we can restrict our focus to the event that $\Pxn, \Pym$ are close to $P$, in the sense that 
\begin{equation}
    \label{equation:AssumedCondition}
    \RE{\Pxn}{P}, \RE{\Pym}{P} < \dm.
\end{equation} 
We henceforth analyze $\RE{\Pxn}{\Pym} \geq c_m$ with this condition, and we will show that for $m$ large enough the probability that both conditions hold is $0$.

Using \Cref{theorem:PinskersInequality}, we can connect our bounds on the KLD to bounds on the total-variation distance between the two empirical distributions. 
\begin{align}
    \dm \geq \RE{\Pxn}{P}, \RE{\Pym}{P} &\geq \frac{1}{2\ln(2)} \oneNorm{\Pxn - P}^2 , \oneNorm{\Pym - P}^2\\
    \implies \oneNorm{\Pxn - P}, \oneNorm{\Pym - P} &\leq \left(2 \ln(2) \dm\right)^{1/2}
\end{align}

From \cite[Equation 7.34]{polyanskiy2025information}, we have 
\begin{align}
    \RE{\Pxn}{\Pym} &\leq \log(e) \cdot \sum_{i=1}^d \frac{(\pxi - \pyi)^2}{\pyi}
\end{align}

% Then since $\ln(x) \leq x - 1$ for $x > 0$, we have 
% \begin{align}
%     \RE{\Pxn}{\Pym} &= \sum_{i=1}^d \pxi \log \frac{\pxi}{\pyi}\\
%     &= \frac{1}{\ln 2} \sum_{i=1}^d \pxi \ln \frac{\pxi}{\pyi}\\
%     &\leq \frac{1}{\ln 2} \sum_{i=1}^d \pxi \frac{\pxi - \pyi}{\pyi}\\
%     &= \frac{1}{\ln 2} \sum_{i=1}^d \frac{(\pxi - \pyi)^2}{\pyi} + (\pxi - \pyi)\\
%     &= \frac{1}{\ln 2} \sum_{i=1}^d \frac{(\pxi - \pyi)^2}{\pyi}
% \end{align}

Now, since $\dm \searrow 0$, $\exists N_{1}>0$ such that $\forall m \geq N_1$ we have $\oneNorm{\Pym - P} \leq \frac{\pmin}{2}$, so that $\pyi \geq \frac{\pmin}{2}$. Then assuming $m \geq N_1$ we have 
\begin{align}
    \RE{\Pxn}{\Pym} &\leq \log(e) \sum_{i=1}^d \frac{(\pxi - \pyi)^2}{\pyi}\\
    &\leq \frac{2\log(e)}{\pmin} \sum_{i = 1}^d (\pxi - \pyi)^2\\
    &\leq \frac{2\log(e)}{\pmin } \oneNorm{\Pxn - \Pym}^2\\
    &\leq \frac{16}{\pmin} \dm
\end{align}

where the last inequality is because 
\begin{align}
    \oneNorm{\Pxn - \Pym}^2 &\leq 2 \oneNorm{\Pxn - P}^2 + 2 \oneNorm{\Pym - P}^2\\
    &\leq 4 \cdot (\dm 2 \ln 2)\\
    &= 8 \ln 2  \cdot \dm
\end{align}

But since $\dm = \Theta(\frac{\log m}{m})$ and $c_m = \Theta(m^{-1/2})$, $\exists N_2 > 0$ such that 
\begin{equation}
    \frac{16}{\pmin} \dm < c_m
\end{equation}

Take $N_{\max} = \max \{N_1, N_2\}$, then for $m \geq N_{\max}$,
\begin{align}
    \RE{\Pxn}{P}, \RE{\Pym}{P} &\leq \dm\\
    \implies \RE{\Pxn}{\Pym} &< c_m
\end{align}

Therefore, for $m \geq N_{\max}$ we have 
\begin{equation}
    \Proba{\RE{\Pxn}{\Pym} \geq c_m | \Pxn, \Pym \in B(P, \dm)} = 0
\end{equation}
which gives us our asymptotic type I error bound.

\subsection{Type II Error}
\label{subsection:TwoSample_TypeIIErrorBound}

When $\supp(P) \cap \supp(Q) = \emptyset$, then we will always have $\RE{\Pxn}{\Qym} = +\infty$, from which we get our first result for the type II error. 

% \vspace{1cm}

Now without loss of generality, we can assume that $\supp(P) \cap \supp(Q) \neq \emptyset$. Take $\tilde w = \frac{n}{n+m}$, so that $\lim_{n,m \to \infty} \tilde w = w$.
Notice that $F^*$ is well defined and unique, this was shown previously in \cite{nielsen2009sided} but for convenience we show it here in \Cref{subsection:OptimalF}. Here we will be using $\Pxn$ referring to the empirical distribution of the samples $X^n$, and $\Qym$ referring to the empirical distribution of the samples $Y^m$. 

Let $K_{n,m} = \{(\Pxn, \Qym) \in \Pcal^d_n \times \Pcal^d_m: \RE{\Pxn}{\Qym} \leq c_m \}$. We have the following argument

\begin{align*}
    \Proba{\hat{H}_0 | H_1} &= \Proba{\RE{\Pxn}{\Qym} < c_m}\\
    &= \sum_{\dot{Q} \in \Pcal^d_m} \ \ \sum_{\dot{P} \in B_n(\dot{Q}, c_m)} \Proba{\Pxn = \dot{P}, \ \ \Qym = \dot{Q}}\\
    &= \sum_{\dot{Q} \in \Pcal^d_m} \ \ \sum_{\dot{P} \in B_n(\dot{Q}, c_m)} \Proba{\Pxn = \dot{P}} \ \ \Proba{\Qym = \dot{Q}}\\
    &\leq \sum_{\dot{Q} \in \Pcal^d_m} \ \ \sum_{\dot{P} \in B_n(\dot{Q}, c_m)} \max_{\substack{ \tilde{Q}_{Y^m} \in \Pcal_m^d\\ \tilde{P} \in B_n(\tilde{Q}, c_m) } } \left\{ \Proba{\Pxn = \tilde{P} } \Proba{\Qym = \tilde{Q}}  \right\}\\
    &\leq (m+1)^d(n+1)^{d} 2^{-(n+m) \min_{\substack{  (\Pxn, \Qym) \in K_{n,m}  }} \left[ \tilde w\RE{\Pxn}{P} + (1-\tilde w)\RE{\Qym}{Q}\right] }
\end{align*}
The third equality is because $X^n$ and $Y^m$ are independent. The fifth inequality follows from the fact that there are at most $(n+1)^d$ and $(m+1)^d$ empirical distributions with denominator $n$ or $m$, and from \Cref{theorem:SpecificTypeBound}.

Now, we can focus on bounding 
\begin{equation}
    (P_n^*, Q_n^*) = \argmin_{ \substack{ \Qym \in \mathcal{P}^d \\ \Pxn \in B(\Qym, c_m)} } \left\{\tilde w \RE{\Pxn}{P} + (1-\tilde w)\RE{\Qym}{Q}\right\}
\end{equation}

% We remind ourselves of our previous assumption, that $\supp(Q) \cap \supp(P) \neq \emptyset$. Define $\tilde{\Pcal}^d \subset \Pcal^d$ such that $\forall H \in \tilde{\Pcal}^d$, $\supp(H) \subset \supp(P) \cap \supp(Q)$. Then for any $H \in \tilde{\Pcal}^d$
% \begin{equation}
%     \RE{Q^*_n}{Q} + \RE{P^*_n}{P} \leq \RE{H}{Q} + \RE{H}{P} < +\infty
% \end{equation}
% So, without loss of generality we can further assume $Q_n^* << Q$, as we would otherwise have $\RE{Q_n^*}{Q} = +\infty$. And by definition we have $P_n^* \in B_n(Q^*_n, c_n)$, so that $P_n^* << Q_n^* << Q$,  and also $\RE{P_n^*}{Q} < +\infty$.

Since we assume that $\supp(P) \cap \supp(Q) \neq \emptyset$, without loss of generality we may assume that $Q_m^* << Q$, and $P_n^* << Q_m^* << Q$. Using the same argument employed in the analysis of the type I error, we must have 
\begin{align}
    \oneNorm{P_n^* - Q_m^*} \leq (2\ln(2) c_m)^{1/2}
\end{align}

Then invoking \Cref{lemma:RETransferInequality} along with the fact that $c_m \searrow 0$ as $m\to \infty$, we get that $\forall \eta > 0$, $\exists N > 0$ such that
\begin{equation}
    \abs{ \RE{P_n^*}{Q} - \RE{Q_m^*}{Q} } \leq \eta, \quad \forall m > N
\end{equation}

Let $F_{\tilde w} \in \Pcal^d$ be such that $f_{\tilde w, i}= \frac{p_i^{\tilde w} q_i^{1-\tilde w}}{\sum_{j=1}^d p_j^{\tilde w} q_j^{1-\tilde w}}$. Then, for $n$ sufficiently large we have
\begin{align}
    \min\limits_{\substack{\Qym \in \Pcal^d \\ \Pxn \in B_n(\Qym, c_m)}} \tilde w\RE{\Pxn}{P} + (1-\tilde w)\RE{\Qym}{Q} &= \tilde w\RE{P_n^*}{P} + (1-\tilde w)\RE{Q_m^*}{Q}\\
    &\geq \tilde w\RE{P_n^*}{P} + (1-\tilde w)\RE{P_n^*}{Q} - (1-\tilde w)\eta\\
    &\geq \tilde w\RE{F_{\tilde w}}{P} + (1-\tilde w)\RE{F_{\tilde w}}{Q} - \eta,
\end{align}
where the last inequality follows because $F_{\tilde w}$ is the minimizer of the expression, as will be shown in \Cref{subsection:OptimalF}.

So we finally get that $\forall \eta > 0$, $\exists N > 0$ such that $\forall m \geq N$: 
\begin{align}
    \Proba{\hat{H}_0 | H_1} &\leq (m+1)^{d}(n+1)^d 2^{-(n+m) \left[\tilde w\RE{F_{\tilde w}}{P} + (1-\tilde w)\RE{F_{\tilde w}}{Q} - \eta\right]}\\
    \implies \liminf_{n,m \to \infty} -\frac{1}{n+m} \log\beta(A_{n,m}) &\geq  w\RE{F^*}{P} + (1- w)\RE{F^*}{Q}
\end{align}

\subsection{Optimal F}
\label{subsection:OptimalF}

We restate the definition of $F^*$ here for the sake of clarity,
\begin{equation}
    F^* = \argmin_{F \in \Pcal^d} w\RE{F}{P} + (1-w)\RE{F}{Q}.
\end{equation}
Here we are allowed to take the minimum over all probability distributions since $\Pcal^d$ is convex and compact, and the function being minimized is convex. We can solve for $F^*$ using Lagrangian optimization. Let $F = (f_1, \dots, f_d)$ be an optimization variable, and $P = (p_1, \dots, p_d)$, $Q = (q_1, \dots, q_d)$ be constants (with respect to the optimization). Imposing the constraint $\sum_{i=1}^d f_i = 1$, then our Lagrangian would be 
\begin{align*}
    L(F, \lambda) &= w\RE{F}{P} + (1-w)\RE{F}{Q} - \lambda \left(1 - \sum_{i=1}^d f_i\right)\\
    &= \sum_{i=1}^d f_i \left(w\log\left(\frac{f_i}{p_i}\right) + (1-w)\log\left(\frac{f_i}{q_i}\right)\right) - \lambda\left(1 - \sum_{i=1}^d f_i\right)\\
    &= \sum_{i=1}^d f_i \log\left(\frac{f_i}{p_i^w q_i^{1-w}}\right) - \lambda \left(1 - \sum_{i=1}^d f_i\right)
\end{align*}

Now, for each $j \in [d]$, we have 
\begin{align*}
    \frac{\partial}{\partial f_j} L(F, \lambda) &= \frac{\partial}{\partial f_j} \left[f_j \log\left(\frac{f_j}{p_j^w q_j^{1-w}}\right)\right] + \lambda\\
    &= \log(f_j) + \log(e) - \log(p_j^w q_j^{1-w}) + \lambda
\end{align*}

Setting this to $0$ yields 
\begin{align*}
    % \log(f_j) &= \log(p_j q_j) - \left(\frac{1}{\ln(2)} + \lambda\right)\\
    f_j &= p_j^w q_j^{1-w} \cdot 2^{-\left(\log(e) + \lambda\right)}
\end{align*}

And the constraint 
\begin{align*}
    \frac{\partial}{\partial \lambda} L(F, \lambda) &= 1 - \sum_{i=1}^d f_j\\
    &= 1 - 2^{-\left(\log(e) + \lambda\right)} \sum_{i=1}^d p_i^w q_i^{1-w}
\end{align*}
is satisfied when $2^{-\left(\log(e) + \lambda\right)} = \sum_{i=1}^d p_i^w q_i^{1-w}$. 

Therefore, our optimal $F^*$ has
\begin{equation*}
    f_j^* = \frac{p_j^w q_j^{1-w}}{\sum_{i=1}^d p_i^w q_i^{1-w}} 
\end{equation*}

\section{Proof of \texorpdfstring{\Cref{theorem:TS_Converse}}{Two-Sample Converse}}
\label{section:Appendix_TS_Converse}

We first need to establish that the empirical distribution $\Pxn$ is a sufficient statistic for any inference on $P$. We have 
\begin{align}
    \Proba{X^n | X^n \sim P} = n! \cdot \prod_{i=1}^d \frac{(P[i])^{n \cdot \Pxn[i]}}{(n \cdot \Pxn[i])!}.
\end{align}
Therefore, by the Neyman-Fisher Factorization Theorem \cite{casella2024statistical}, we know that $\Pxn$ is a sufficient statistic for any inference on $P$. From this we can see that $(\Pxn, \Qym)$ is a sufficient statistic for the two-sample problem, and following standard sufficiency arguments \cite{lehmann2005testing} without loss of generality we may restrict our attention to tests $T_{n,m} \subset \Pcal^d_n \times \Pcal^d_m$.

The next result shows that for a test satisfying the type I error condition of \Cref{theorem:TS_Converse}, it must have a non-empty intersection with neighborhoods around the set (simplex) $P=Q$ in the doubled simplex space $\Pcal^d \times \Pcal^d$.

For $F \in \Pcal^d$, for $\delta > 0$, define 
\begin{align*}
    N^2_\delta(F) &:= \{(P,Q) \in \Pcal^d\times \Pcal^d : \oneNorm{P - F}, \oneNorm{Q-F} < \delta\}\\
    G(F) &:= \{(P,Q) \in \Pcal^d\times \Pcal^d: P,Q <<F\}.
\end{align*}
\begin{lemma}[Neighborhood intersections]
    \label{lemma:GoodNeighborhoods}
    Take $T_{n,m}$ a sequence of tests with $T_{n,m} \subset \Pcal_n^d \times \Pcal_m^d$ and $\exists \eps \in (0,1)$ such that $\forall F \in \Pcal^d$
    \begin{equation}
        \limsup_{n,m \to \infty} \Prob[(\Pxn, \Pym) \in T_{n,m}^C | X^n, Y^m \sim F] \leq \eps.
    \end{equation}

    Then $\forall \delta > 0$, and $\forall F \in \Pcal^d$, for $n,m$ large enough, there is some 
    $(\tilde P_n, \tilde Q_m) \in T_{n,m} \cap N_\delta^2 \cap G(F)$. 
\end{lemma}
\begin{proof}[Proof of \Cref{lemma:GoodNeighborhoods}]
    Here we will fix some arbitrary $F \in \Pcal^d$, and assume that $X^n, Y^m \sim F$. With this assumption, 
    \begin{equation*}
        \Proba{(\Pxn, \Qym) \in T_{n,m}} = \Proba{(\Pxn, \Qym) \in T_{n,m} \cap G(F)}.
    \end{equation*}
    Now, since $\Pxn, \Qym \to F$ \cite{van2000asymptotic}, for $\eps < \eta < 1$, when $n,m$ are large enough we have 
    \[
        \Proba{(\Pxn, \Qym) \in N_\delta^2(F)} \geq \eta.
    \]

    By the premise of the Lemma, for $n,m$ large enough,
    \[
        \Proba{(\Pxn, \Qym) \in T_{n,m}} \geq 1-\eps.
    \]
    Therefore, taking $n,m$ larger than in the above conditions, we get 
    \begin{align*}
        \Proba{(\Pxn, \Qym) \in T_{n,m} \cap G(F) \cap N_\delta^2(F)} &\geq \Proba{(\Pxn, \Qym) \in T_{n,m} \cap G(F)} + \Proba{(\Pxn, \Qym) \in N_\delta^2(F)} -1\\
        &\geq \eta - \eps > 0.
    \end{align*}
\end{proof}

Following the result of \Cref{lemma:GoodNeighborhoods}, for any $F \in \Pcal^d$ we can find a sequence of empirical distributions, $\{\tilde F_n, \tilde F_m\}$, inside $T_{n,m}$ that are well behaved and such that $\tilde F_n, \tilde F_m \to F$. 
\begin{corollary}[Well behaved convergent sequence in $T_{n,m}$]
    \label{corollary:wellBehavedConv}
    For $T_{n,m}$ a sequence of tests satisfying the condition of \Cref{lemma:GoodNeighborhoods}, then for any $F \in \Pcal^d$, $\exists \{\tilde F_n, \tilde F_m\}$ so that 
    \begin{align*}
        (\tilde F_n, \tilde F_m) &\in T_{n,m}, \quad \forall n,m\\
        \tilde F_n, \tilde F_m &\to F\\
        \tilde F_n, \tilde F_m &<< F, \quad \text{eventually}
    \end{align*}    
\end{corollary}
\begin{proof}[Proof of \Cref{corollary:wellBehavedConv}]
    For $k \in \N$, let $\delta_k = \frac{1}{k}$, and from \Cref{lemma:GoodNeighborhoods}, take $N_k > 0$ such that $\forall n,m > N_k$ there is $(\tilde P_n, \tilde Q_m)_k \in T_{n,m} \cap N_\delta^2(F) \cap G(F)$. Without loss of generality we may assume that the $N_k$ are strictly increasing with $k$, so that $N_1 < N_2 < \dots$. 
    
    For $\min\{n,m\} < N_1$, choose $\tilde F_n, \tilde F_m$ arbitrarily from $T_{n,m}$. For $n \geq N_1$, take 
    \[
        k_{n,m} := \max\{k : N_k \leq \min(n,m)\},
    \]
    and select
    \[
        (\tilde F_n, \tilde F_m) := (\tilde P_n, \tilde Q_m)_{k_{n,m}}.
    \]
    Notice that $k_{n,m} \to \infty$ as $n,m \to \infty$, and $\forall n,m \geq N_1$ we have $\tilde F_n, \tilde F_m << F$, proving the result. 
\end{proof}

To prove the converse, we will use the following fact about the KLD \cite[Proposition4.8]{polyanskiy2025information}.
\begin{proposition}[KLD is continuous when finite]
    \label{proposition:KLDContinuous}
    On $C(Q) = \{P \in \Pcal^d: P << Q\}$, $D(P\|Q)$ is continuous. 
\end{proposition}

Now, we can prove the converse. 
\begin{proof}[Proof of \Cref{theorem:TS_Converse}]
    Fix some $P,Q \in \Pcal^d$ with $P \neq Q$ and $\supp(P) \cap \supp(Q) \neq \emptyset$, and take $X^n \sim P$ and $Y^m \sim Q$. 
    
    Define $F_w \in \Pcal^d$ so that 
    \[
        f_{w,i} = \frac{p_i^w q_i^{1-w}}{\sum_{j=1}^d p_j^w q_j^{1-w}}.
    \]
    Take $\tilde F_n, \tilde F_m$ from \Cref{corollary:wellBehavedConv}, so that $\tilde F_n, \tilde F_m \to F_w$. We also have $\tilde F_n, \tilde F_m << F_w$ eventually, so that $\tilde F_n << F_w << P$ and $\tilde F_m << F_w << Q$.

    Take $\tilde w = \frac{n}{n+m}$, so that $\tilde w \to w$ by assumption. From the lower bound in \Cref{theorem:SpecificTypeBound}, 
    \begin{align}
        \Proba{(\Pxn, \Qym) \in T_{n,m}} &\geq \Proba{(\Pxn, \Qym)  = (\tilde F_n, \tilde F_m)}\\
        &\geq ((n+1)(m+1))^{-d} 2^{-(n+m) [\tilde w D(\tilde F_n\|P) + (1-\tilde w) D(\tilde F_m\|Q)]}.
    \end{align}
    So that 
    \begin{equation}
        \limsup_{n,m \to \infty} - \frac{1}{n+m} \log \Proba{(\Pxn, \Qym) \in T_{n,m}} \leq \limsup_{n,m\to \infty} \tilde w D(\tilde F_n\|P) + (1-\tilde w) D(\tilde F_m\|Q). 
    \end{equation}
    Now, since $\tilde F_n << P$ and $\tilde F_m << Q$ eventually, using \Cref{proposition:KLDContinuous}, we have
    \begin{align}
        \limsup_{n,m\to \infty } \tilde w D(\tilde F_n\|P) + (1-\tilde w) D(\tilde F_m\|Q) &= \lim_{n,m \to \infty} \tilde w D(\tilde F_n\|P) + (1-\tilde w) D(\tilde F_m\|Q). \\
        &= w D(F_w\|P) + (1-w) D(F_w\|Q).
    \end{align}
    And from \Cref{subsection:OptimalF}, we have the result. 
\end{proof}

\section{Proof of \texorpdfstring{\Cref{theorem:TS_StrongConverseRate}}{Two-Sample Strong Converse Rate}}
\label{section:Appendix_TS_StrongConverseRate}

Throughout, we will use $P,Q$ to denote the fixed pair from the theorem statement, and take $F \equiv F^*$. Let 
\begin{align*}
    K(\delta) &:= \{(A,B) \in \Pcal^d \times \Pcal^d : w D(A\|P) + (1-w) D(B\|Q) \geq w D(F\|P) + (1-w) D(F\|Q) + \delta\},\\
    \Gamma(\delta) &:= \inf\{w D(A\|F) + (1-w) D(B\|F): (A,B) \in K(\delta)\}.
\end{align*}

We start by showing that for any $0< \eta <\delta$, $\Gamma(\eta)$ is a lower bound on the exponent of getting the null hypothesis correct.

\begin{lemma}
    \label{lemma:TS_StrongConverse_GammaDelta}
    Assume the same conditions as in \Cref{theorem:TS_StrongConverseRate}, then we have, for any $0 < \eta < \delta$,  
    \begin{equation}
        \liminf_{n,m \to \infty} -\frac{1}{n+m}\log \Proba{(\Pxn, \Qym) \in T_{n,m} | X^n, Y^m \sim F} \geq \Gamma(\eta).
    \end{equation}
\end{lemma}
\begin{proof}
    Take $\tilde w = \frac{n}{n+m}$. From the lower bound in \Cref{theorem:SpecificTypeBound}, and the type II error condition, one can verify that for $n,m$ large enough, $T_{n,m} \subset K(\eta)$. 
    Therefore for $n,m$ large enough, using \Cref{theorem:SanovsTheorem} and assuming $X^n, Y^m \sim F$, we get 
    \begin{align*}
        \Proba{(\Pxn, \Qym) \in T_{n,m}} &\leq (n+1)^d(m+1)^d 2^{-(n+m) \inf\limits_{(A,B) \in T_{n,m}} \tilde w D(A\|F) + (1-\tilde w) D(B\|F) }\\
        &\leq (n+1)^d (m+1)^d 2^{-(n+m) \inf\limits_{(A,B) \in K(\eta)} w D(A\|F) + (1-w) D(B\|F)}\\
        &= (n+1)^d (m+1)^d 2^{-(n+m) \Gamma(\eta)}.
    \end{align*}
\end{proof}

Now we can show the theorem. 
\begin{proof}[Proof of \Cref{theorem:TS_StrongConverseRate}]
    Thanks to \Cref{lemma:TS_StrongConverse_GammaDelta}, we only need to show that 
    \[
        \Gamma(\eta) \geq \sup_{0 \leq s< 1-w} \left\{s\eta+ (1-s) \left[\psi(w) - w \psi\left(\frac{w-s}{1-s}\right) - (1-w) \psi\left(\frac{w}{1-s}\right)\right] \right\}.
    \]

    Take 
    \begin{align*}
        J_w(A,B) &:= w D(A\|F) + (1-w) D(B\|F)\\
        T_w(A,B) &:= w D(A\|P) + (1-w) D(B\|Q) - J_w(A,B) - (1-w) D_w(P\|Q)\\
        &= w(1-w) \sum_{i=1}^d b_i \log \frac{p_i}{q_i} - w(1-w) \sum_{i=1}^d a_i \log \frac{p_i}{q_i}.
    \end{align*}
    Using \Cref{equation:RenyiRel_MidPointInterp}, one can verify that 
    \begin{align*}
        \Gamma(\eta) &= \inf\{wD(A\|F) + (1-w)D(B\|F) : A,B \in K(\eta)\}\\
        &= \inf\{J_w(A,B) : J_w(A,B) + T_w(A,B) \geq \eta; A,B \in \Pcal^d\}.
    \end{align*}

    We introduce a slack term to simplify the optimization problem. For any $s \in [0,1-w)$, and any viable $A,B$, we have 
    \begin{align*}
        J_w(A,B) &= (1-s) J_w(A,B) + s J_w(A,B)\\
        &\geq (1-s) J_w(A,B) + s(\eta - T_w(A,B)).
    \end{align*}
    Therefore, $\forall 0 \leq s < 1-w$, we have 
    \begin{equation*}
        \Gamma(\eta) \geq s\eta + \inf_{A,B}\{ (1-s) J_w(A,B) - s T_w(A,B)\}
    \end{equation*}

    Fixing some $s \in [0,1-w)$, then 
    \begin{align*}
        (1-s) J_w(A,B) - s T_w(A,B) &= (1-s) w D(A\|F) + (1-s) (1-w) D(B\|F) \\
        &\ \ \ - s w(1-w) \sum_{i} b_i \log \frac{p_i}{q_i} + s w (1-w) \sum_i a_i \log \frac{p_i}{q_i}\\
        &= G_1(A) + G_2(B),
    \end{align*}
    where 
    \begin{align*}
        G_1(A) &:= (1-s) w D(A\|F) + s w (1-w) \sum_{i}a_i \log \frac{p_i}{q_i}\\
        G_2(B) &:= (1-s) (1-w) D(B\|F) - s w (1-w) \sum_{i}b_i \log \frac{p_i}{q_i}
    \end{align*}
    Therefore, 
    \begin{align*}
        \Gamma(\eta) &\geq s \eta + \inf_A  G_1(A) + \inf_BG_2(B) 
    \end{align*}

    Now, taking $z_1 = (1-s) w$, $\ell_i = -s w (1-w) \log \frac{p_i}{q_i}$, we have 
    \begin{align*}
        G_1(A) &= z_1 \left[D(A\|F) - \sum_{i=1}^d a_i \frac{\ell_i}{z_1}  \right]\\
        \implies \inf_{A \in \Pcal^d} G_1(A) &= - z_1 \log \left(\sum_{i=1}^d f_i 2^{\frac{\ell_i}{z_1}}\right)\\
        &= z_1 \psi(w) - z_1\log \left(\sum_{i=1}^d p_i^{\frac{w-s}{1-s}} q_i^{1 - \frac{w-s}{1-s}}\right)\\
        &= (1-s)w \left[\psi(w) - \psi\left(\frac{w-s}{1-s}\right)\right].
    \end{align*}
    Here, we arrive at the second equality using \Cref{theorem:GibbsVariationalPrinciple}.

    And taking $z_2 := (1-s) (1-w)$, $k_i := s w(1-w) \log \frac{p_i}{q_i}$, we get 
    \begin{align*}
        G_2(B) &= z_2 \left[ D(B\|F) - \sum_{i=1}^d b_i \frac{k_i}{z_2}\right]\\
        \implies \inf_{B \in \Pcal^d} G_2(B) &= - z_2 \log\left(\sum_{i=1}^d f_i 2^{\frac{k_i}{z_2}}\right)\\
        &= z_2 \psi(w) - z_2 \log \left(\sum_{i=1}^d p_i^{\frac{w}{1-s}} q_i^{1- \frac{w}{1-s}}\right)\\
        &= (1-s) (1-w) \left[\psi(w) - \psi\left(\frac{w}{1-s}\right)\right].
    \end{align*}

    Then in combination we have 
    \begin{align*}
        \Gamma(\eta) &\geq s \eta + \inf_{A} G_1(A) + \inf_B G_2(B)\\
        &= s \eta + (1-s) w \left[\psi(w) - \psi\left(\frac{w-s}{1-s}\right)\right] + (1-s) (1-w) \left[\psi(w) - \psi\left(\frac{w}{1-s}\right)\right]\\
        &= s\eta + (1-s) \left[\psi(w) - w \psi\left(\frac{w-s}{1-s}\right) - (1-w) \psi\left(\frac{w}{1-s}\right)\right]
    \end{align*}
    And since this holds for all $s \in [0,1-w)$, we have
    \begin{equation*}
        \Gamma(\eta) \geq \sup_{0\leq s  < 1-w} \left\{ s\eta + (1-s) \left[\psi(w) - w \psi\left(\frac{w-s}{1-s}\right) - (1-w) \psi\left(\frac{w}{1-s}\right)\right] \right\}.
    \end{equation*}

    The last step is to show that this is strictly positive, for $\eta > 0$. To see this, take 
    \[
    R_\eta(s) := s \eta +(1-s) \left[\psi(w) - w \psi\left(\frac{w - s}{1-s}\right) - (1-w) \psi\left(\frac{w}{1-s}\right)\right].
    \]
    Notice that 
    \begin{align*}
        R_\eta(0) &= 0\\
        \frac{d}{ds} R_\eta(0) &= \eta.
    \end{align*}
    Therefore, $\exists s \in (0,1-w)$ such that $R_\eta > 0$. And since we have $\forall s \in (0,1-w)$, $\Gamma(\eta) \geq R_\eta(s)$, the result follows. 
\end{proof}

\fi

\end{document}